# THE DETERMINATION OF THE WATER VAPOR CONTENT IN THE PULKOVO VKM-100 MULTIPASS VACUUM CELL USING POLYMER SENSORS OF HUMIDITY.

**Galkin V. D.[1), Naebert T.[2), Nikanorova I. N.[1), Sal'nikov I. B.[1), Leiterer U.[2), Alekseeva G. A.[1), Novikov V. V.[1), and Dauß D.[2).**

*1)* The Central (Pulkovo) Astronomical Observatory, Russian Academy of Sciences, Russia
*2)* Deutscher Wetterdienst, Meteorologisches Observatorium Lindenberg, Germany

*In spectral studies of water vapor under laboratory conditions (determination of molecular constants, measurement for spectral transmission functions), the amount of water vapor in the time of the measurements is one of the most essential parameters, which should be determined accurately. We discuss the application for this purpose of polymer sensors of humidity manufactured by Praktik-NC (Moscow) and used in the Pulkovo VKM-100 multipass vacuum cell. These sensors were examined in the laboratory of Lindenberg Meteorological observatory (Germany) by comparison between their readings and those of standard measuring devices for various values of relative humidity, pressure, and temperature. We also carried out measurements of relative humidity in boxes with saline solution, in which the relative humidity that corresponds to a given solution is guaranteed with the accuracy of several tenths of percent. The analysis of the results of the laboratory examination of the sensors and extended sets of measurements made with the Pulkovo cell made it possible to conclude that in measurements in the interval of relative humidity 40-80%, the ~5% accuracy of the measurements for the water vapor content is reached. Further paths are indicated for the increase of the accuracy of measurements and extending the interval of the relative humidity, in which accurate measurements may be carried out.*

## Introduction.

Solar and stellar photometers used in studies for optical parameters of the atmosphere, may be also used to measure atmospheric water vapor content. In this case, one of the filters of the photometer should be centered on the absorption band of the water vapor. In observations in a water band, the reading of the photometer will depend not only on the optical characteristics of the atmosphere at a given wavelength, but also on the amount of water vapor along the line of view. The technique used in the observations and data processing [1, 2] makes it possible to discriminate the fraction of the signal attenuation related to the absorption by water vapor. If the corresponding calibration dependence for the absorption as a function of the water vapor content along the line of view is available, the atmospheric water vapor content may be derived from observations. Currently, the most accurate way to obtain this calibration dependence is the direct calibration of photometrical data obtained under laboratory conditions as a function of the number of absorbing molecules along the line of view. To this end, in Pulkovo Observatory the purpose-made laboratory complex is used, which we described previously [3]. The VKM-100 cell with multiple light passages makes it possible to vary the number of absorbing molecules along the line of view by a known factor, due to the variation of the number of passages of light in the cell. With the mirrors currently used in the cell, the path length may vary from 100 m to 4100 m, i.e. the number of absorbing molecules along the line of view may be varied by the factor of 41. However, the number of absorbing molecules that account for one passage remains unknown. If our calibration should provide the accuracy of 1-2% in the atmospheric determination of the water vapor content with a photometer, at least not a lower accuracy of determination of the water vapor content in the cell should be provided in the process of the photometer calibration. This metrological problem is far from simple. Currently, it is not possible to mount standard means of humidity measurements in the cell. Therefore, it was decided to use polymer sensors of relative humidity produced for technical purposes, and to carry out their testing and calibration under laboratory conditions independently. Since a sensor provides local information on humidity, several sensors were placed along the cell, and their readings were compared and analyzed for various conditions of observations. The sensors were tested in the PTU Laboratory (Pressure, Temperature, and Humidity) at Lindenberg Meteorological observatory (Germany). It was particularly important, since it made it possible not only to study the sensors, but also to compare directly the humidity scales of the Pulkovo sensors and the scale with the reference to which sensors for radiosondes are calibrated. Here, we present the results of the laboratory testing of the sensors and analyze measurements made with these sensors in the cell.

## The device for humidity and temperature measurements

The IVTM-7 MK-C device for humidity and temperature measurements, used in our studies and manufactured by Praktik-NC, Moscow, is assigned for permanent measurements of temperature and humidity in different technological processes in manufacturing industry, power engineering, medicine, and agriculture. The device consists of the primary converter of the relative humidity and temperature IPVT-03-07 (sensor) and the control unit [4].

The basic parameters of the device are as follows: the interval of humidity measurements 2 to 98%, the error of the measurements of the relative humidity 2% (over the total interval of measurements), the supplementary error caused by variations of the surrounding air temperature 0.2% per 1 degree. The error of the temperature measurements in the interval 0 to 40$^{o}$C is 0.5$^{o}$.

The IPVT-03-06 converter is made in accordance with the schema for an RC generator with a timer type 555. As the R-element in the temperature channel, a platinum thermal resistor is used, while as the C-element in the humidity channel, a capacitive sensor of humidity is applied. The converter consists of the aluminum frame, in which the transformation circuit for the signals from humidity and temperature sensors is placed, and the metal probe, in the edge of which, in the measuring cap, the humidity and temperature sensors are located.

The sensors are connected to the timer via electronic switchboard. In addition to the measuring elements, the switchboard connects the timer to standard RC-elements (as standard elements, thermally stable resistors and condensers are used). This schema makes it possible to compensate the converter automatically for variations of the temperature of the surrounding medium.

The control of the switchboard, calculation of the timer frequency, temperature, and humidity is carried out by the logical unit of the converter, which is based on a PIC-controller. A microcontroller code guides the measurement of frequency from the sensors, the functioning standard elements, and the calculation of the temperature and humidity using individual calibrations from the memory of the calculation unit of the converter. The calculated humidity and temperature, in the form of a subsequent digital code, are sent to the output unit of the converter. The unit transmits the measured data to the device via the semi-duplex channel by differential method. This way of transmission makes it possible for the converter to operate over long lines of connection in the presence of high levels of electromagnetic noise. The converter works steadily at the distance up to no less than 300 meters.

The control unit is a microprocessing system based on the Intel MCS-51 microcontroller, which is essentially a microcomputer for operations with the IPVT-03 humidity and temperature converter and its modifications. The control unit is operated by the code recorded in a permanent storage device. Inner variables, calibration constants, and other operation parameters, are stored in the flash memory, which is power-independent and maintains information with disconnected power during the total term of service of the device.

## The location of the IPVT-03-06 sensors in the cell.

Currently, the measuring complex consists of the VKM-100 cell, the light source, the solar photometer offered by Lindenberg Meteorological observatory, and 4 IPTV-03-06 sensors (№№ 65132, 1229, 1230, 1087). **Fig. 1** presents the general view of the complex; F**ig. 2a** – the schema of the location of the sensors in the cell. The sensors are connected to the cell through special mounting flanges and are jointed to the control unit by the cable, the length of which is respectively 5 m for the first sensor, 20 m for the second, 60 m for the third, and 100 m for the fourth. The cables from the sensors are connected to the control unit via a packet-type switch, which read out information from the sensors sequentially. For each sensor, the temperature T, relative humidity H(%), and absolute humidity a (g/m$^3$) are recorded. The average value for the four sensors describes the content of the water vapor in the cell for a given pressure, while the dispersion of individual readings of the sensors yields the error of the measurement. A common series of the measurements for the calibration of photometers started at the atmospheric pressure, when the sensors could be mounted in the cell; if necessary, liquid water was added to increase humidity. Then the air was evacuating step by step to reach sequentially the required pressure of the value, at which the measurements with the photometer were carried out, and the relative and absolute humidity were measured. **Fig. 2b** displays the results of the measurements of the relative humidity as a function of pressure for one of the measurement series.

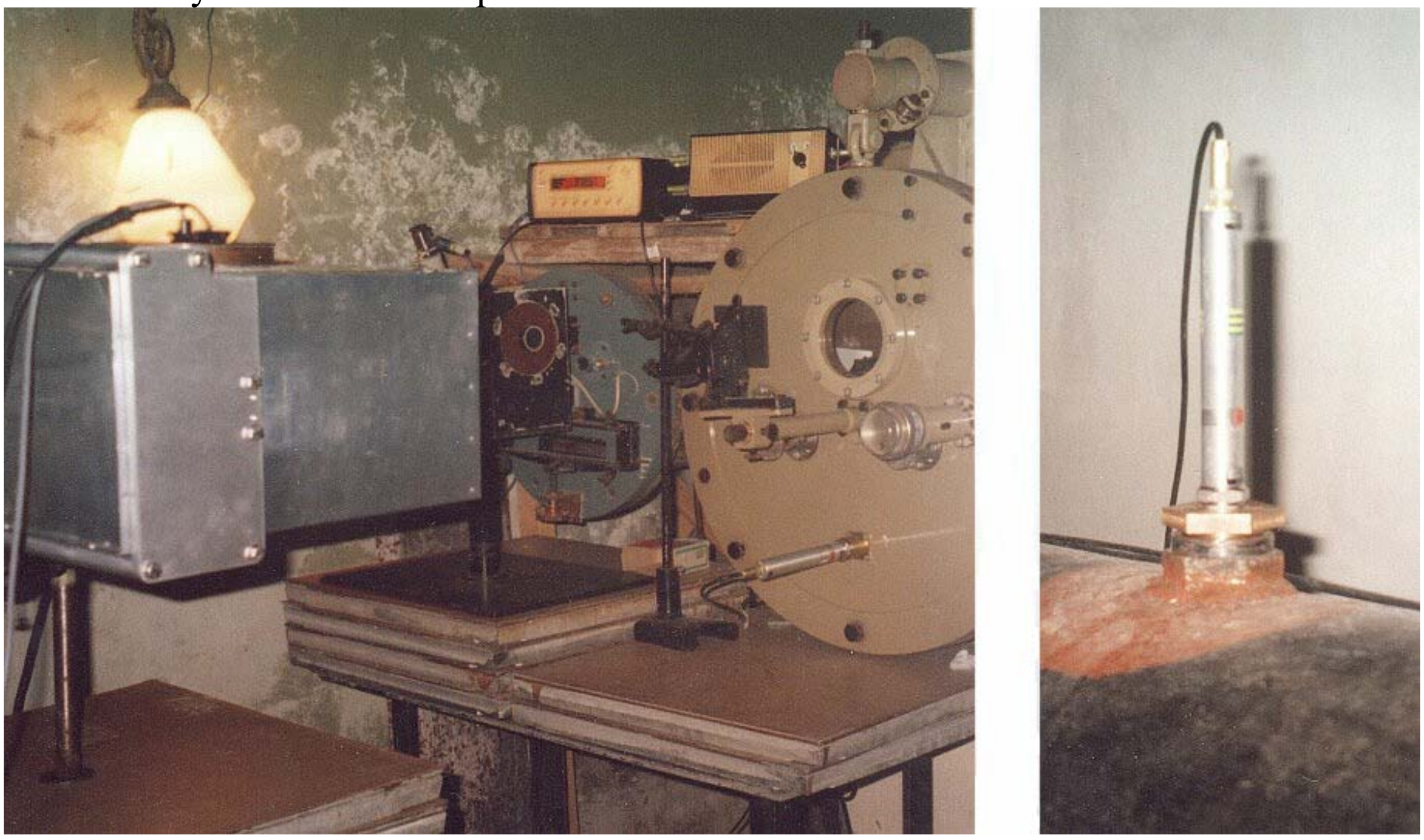

**Fig. 1**. The general view of the measuring complex (solar photometer, control unit with the switchboard, IPVT-03-06 sensors at the entrance and at the distance of 60 m in the VKM-100 cell).

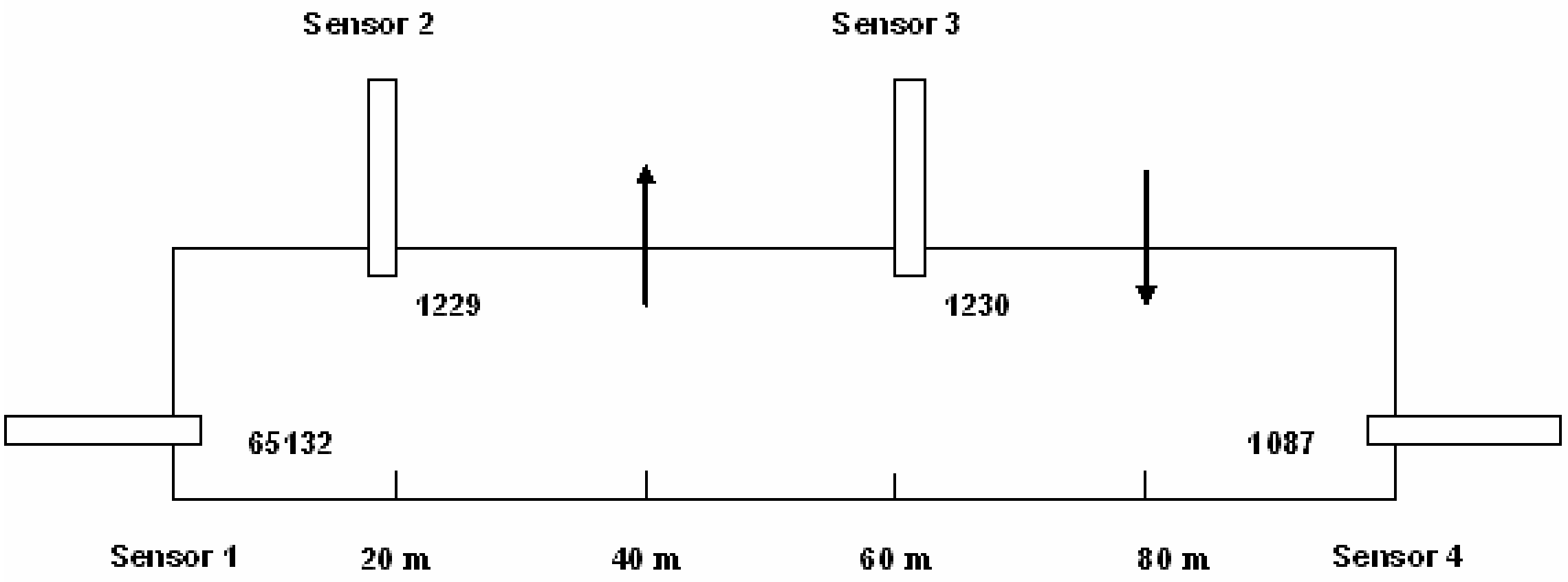

**Fig. 2a**. The location of the sensors in the VKM-100 cell.

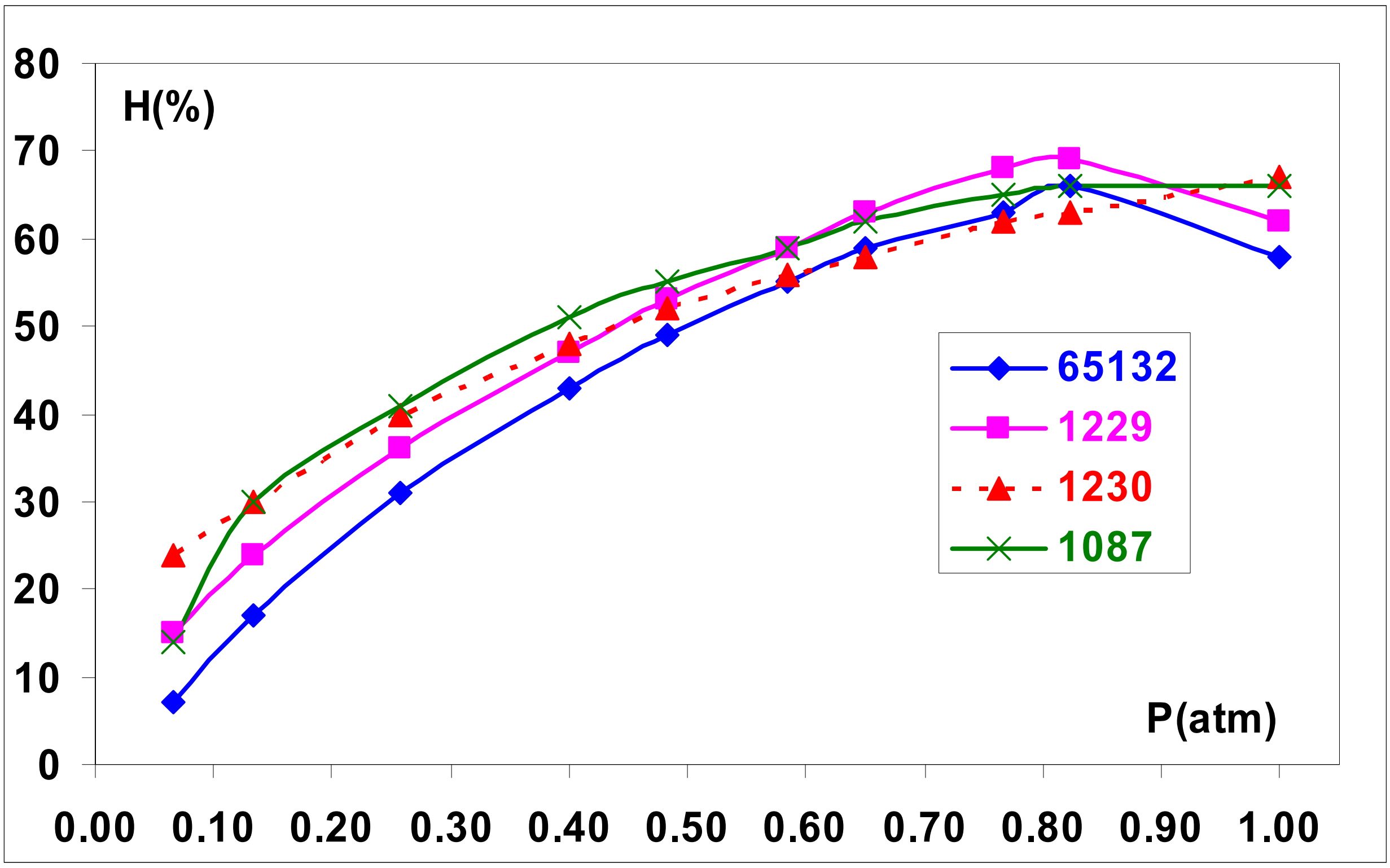

**Fig.2b**. The relative humidity H(%) measured with four sensors as a function of the pressure in the cell.

**Figure 2b** displays the variations of the relative humidity with that of the pressure, and also the degree of consistency between the variations in readings of individual sensors. **Figure 3** presents the same data in the form of the ratio of the measured relative humidity and the average of the measurements of the four sensors for a given pressure, which makes the degree of the deviations of individual measurements from the average more clearly visible.

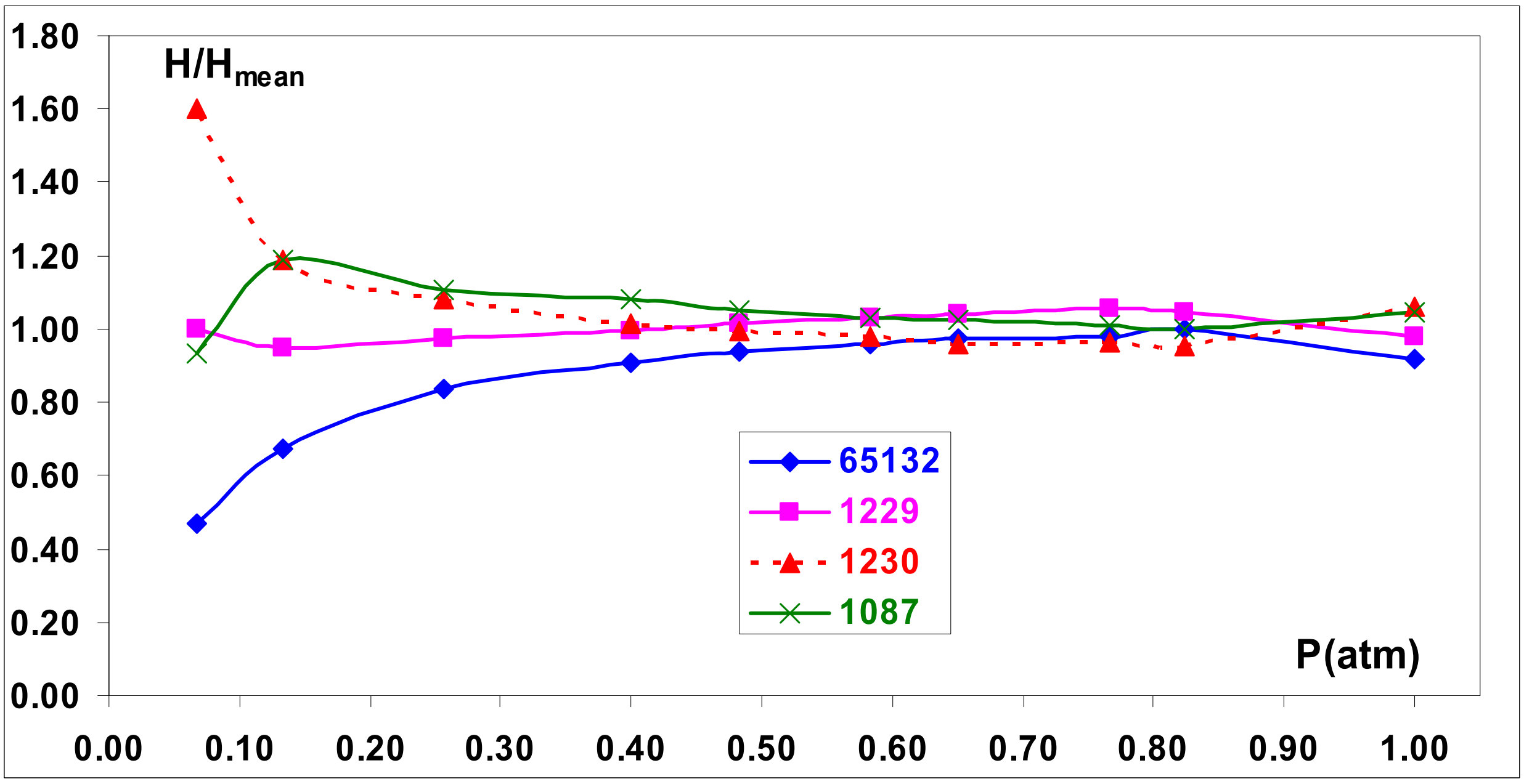

**Fig. 3.** The humidity H**(%)** relative to the average value of H**(%)** for a given pressure.

It follows from **Figs 2-3** that the convergence of individual measurements deteriorates noticeably with the decrease of the pressure and, respectively, with the decrease of the relative humidity.

In order to illustrate variations in the readings of the sensors as a function of their location in the cell, **Fig. 4** presents the readings of individual sensors relative to their average values depending on the location of the sensor in the cell.

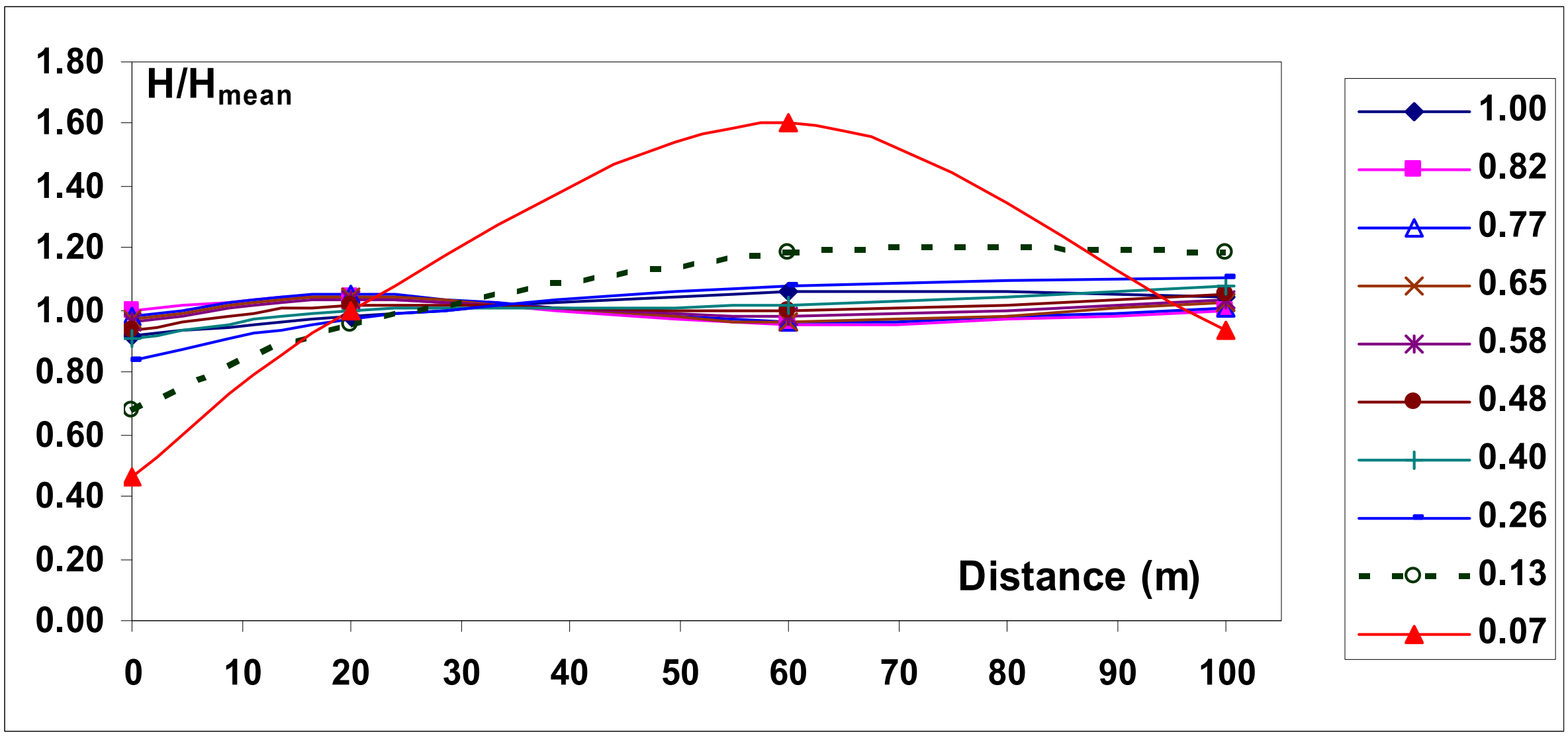

**Fig.4**. The distribution of H(%) in the cell relative to its average value for different pressure values.

**Figures 3** and **4** clearly indicate that the results of the measurements for different sensors are sufficiently compatible (with the r.m.s. deviation ~5%), down to the pressure 0.4 atm; for lower pressure, the error rapidly increases with the decrease of the pressure. As an example, **Figs 2-4** present the case of the summertime observations in the cell, when the heating is off and the temperature and humidity distributions in the cell are uniform. In a similar way, the other cases were analyzed, including those when different values of temperature gradient (up to $15^{o}C$) are established in the cell, and liquid water is introduced to it to increase the humidity.

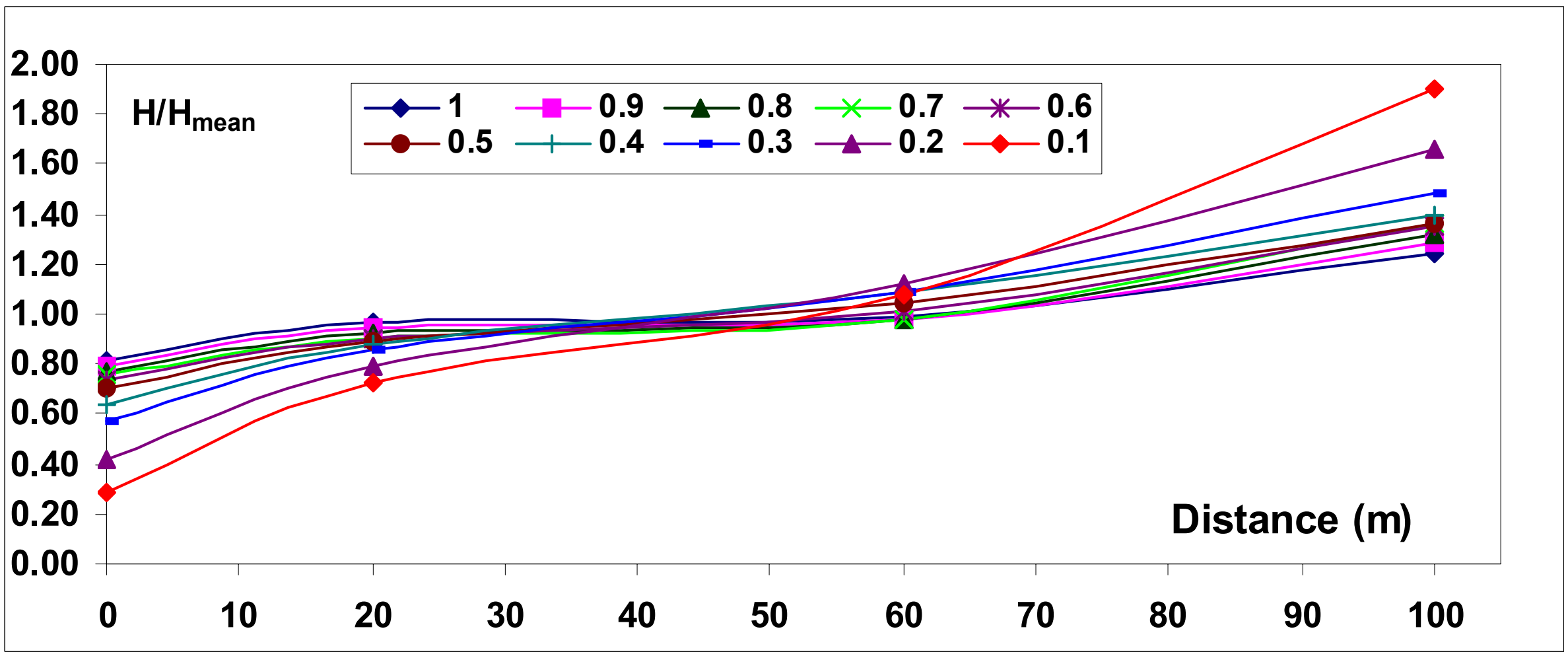

**Fig.5.** The H(%) distribution relative to the average value of H ($H_{mean}$) for various pressure values, in the presence of a temperature gradient along the cell.

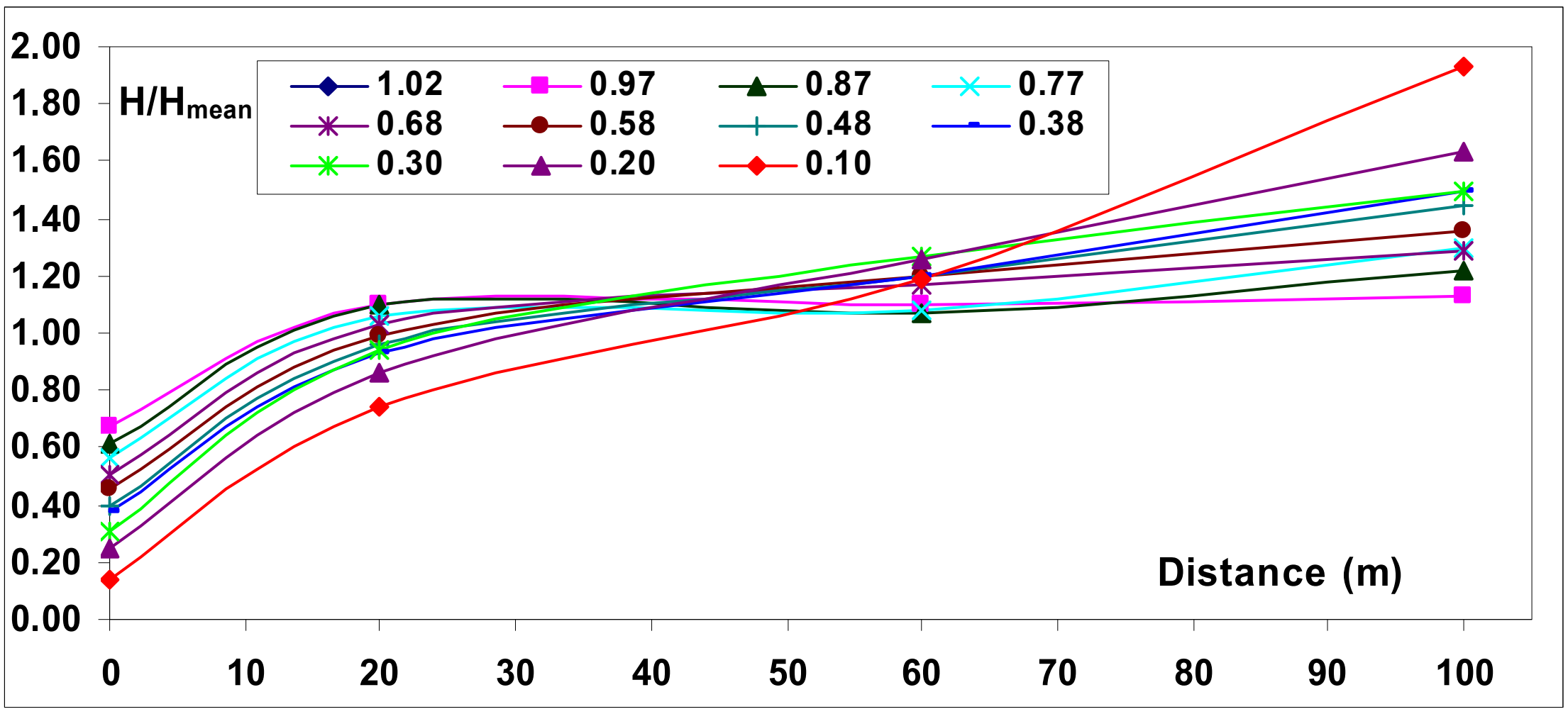

**Fig. 6.** The H(%) distribution relative to the average value of H ($H_{mean}$) for various pressure values, in the presence of a temperature gradient and liquid water in the cell.

**Figure 5** presents the data for the case of a temperature gradient in the cell, while F**ig. 6** reflects the situation when liquid water is introduced to the cell. The latter operation substantially alters the humidity distribution in the cell; however, the established distribution is maintained with various pressure values, down to 0.4 atm (until the humidity falls below 30-40%). According to Figs. 4-6, with the relative humidity above 40% and a uniform temperature distribution inside the cell, the humidity distribution in the cell, normalized to its average value, is maintained at the level of 3%, in the presence of a temperature gradient of 5%, and in the case of introduction of liquid water of 9%. Following this schema, all the measurements of the absolute humidity were considered.

## Testing of Pulkovo sensors at the laboratory of Lindenberg Meteorological observatory (Germany).

In order to carry out independent expert evaluation of the sensors, to estimate their measurement errors, operation stability, individual differences, and to compare the humidity scales, we calibrated the sensors in the laboratory of Lindenberg Meteorological observatory (Germany), which is able to arrange various humidity, temperature, and pressure in isolated volumes (saline boxes and barocamera) and has at its disposal equipment for measurements and testing of the parameters of these conditions.

In 2004, a trial testing of the sensors was carried out, in order to confirm fundamental possibility for their application in the measurements of humidity in the cell.

In 2005 and 2006, in Lindenberg laboratory, measurements made there with the Pulkovo sensors were compared with those carried out with standard measuring equipment. This short-term study was nonetheless more detailed than the previous, and included measurements in boxes above saline solutions (these boxes provide the accuracy of reproduction of relative humidity of several tenths of percent [5]) and measurements of humidity in a barocamera with a fixed pressure (200, 400, 600, 800, and 1000 hPa).

In 2006, this procedure was applied to all Pulkovo sensors (№№ 65132, 1229, 1230 and 1087) with the temperature 10°C and 20°C (in 2005, the measurements were carried out for all sensors with the temperature 10°C and selectively with 20°C). In all conditions, the measurements were carried out with both our sensors and standard means of measuring – the TOROS condensation hygrometer manufactured by COMET (Moscow) and Vaisala sensors (produced in Finland). The processing of the measurements made with the Vaisala sensors with the use of frequency standardization technique custom-designed at Lindenberg laboratory (FN-technique [6]) provides the accuracy of 1% over the total interval of the relative humidity from 0 to 100%, i.e. the same provided by the TOROS condensation hygrometer. Therefore, all measurements made with the Pulkovo sensors were compared to readings of FN equipment, TOROS, and saline solutions, taken as standard measuring means.

Below, we present some results that characterize our sensors. **Figure 7a** displays an example of a set of calibration measurements in the barocamera made on 11.07.2006 in Lindenberg. The upper part of the Figure presents temperature measurements made with the Vaisala sensor according to FN-technique, with the TOROS hygrometer, and with a Pulkovo sensor for 6 min; the lower part shows relative humidity measurements with the same sensors. **Figure 7b** presents the data of the measurements made with all sensors at different conditions (above saline solutions and in the barocamera, with various temperature and pressure values), with the reference to the ideal parameters of the medium (the humidity calculated for saline solutions, the data for the barocamera derived from measurements with the standard TOROS device).

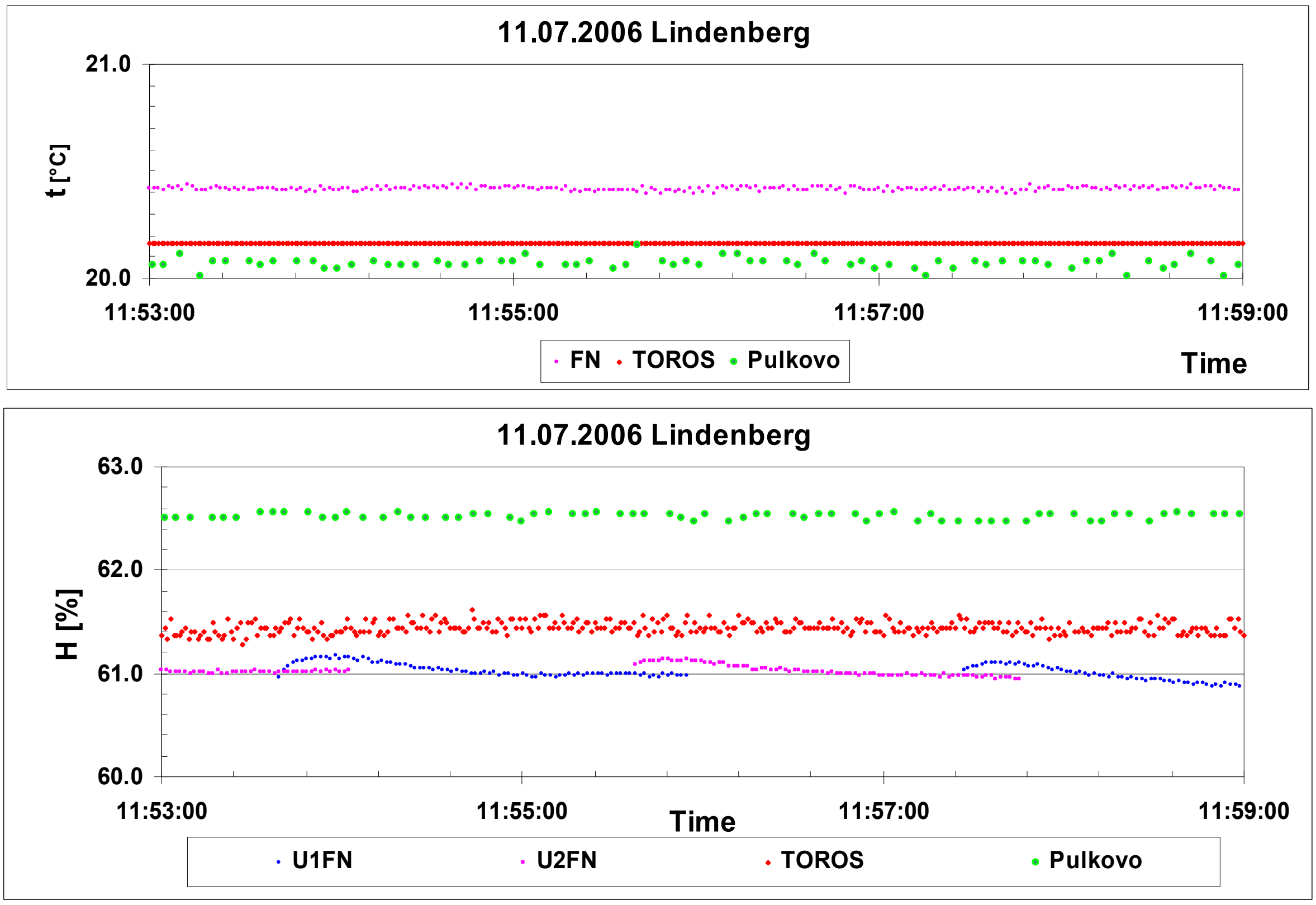

**Fig. 7 a**. An example of a set of calibration measurements made in Lindenberg (sample).

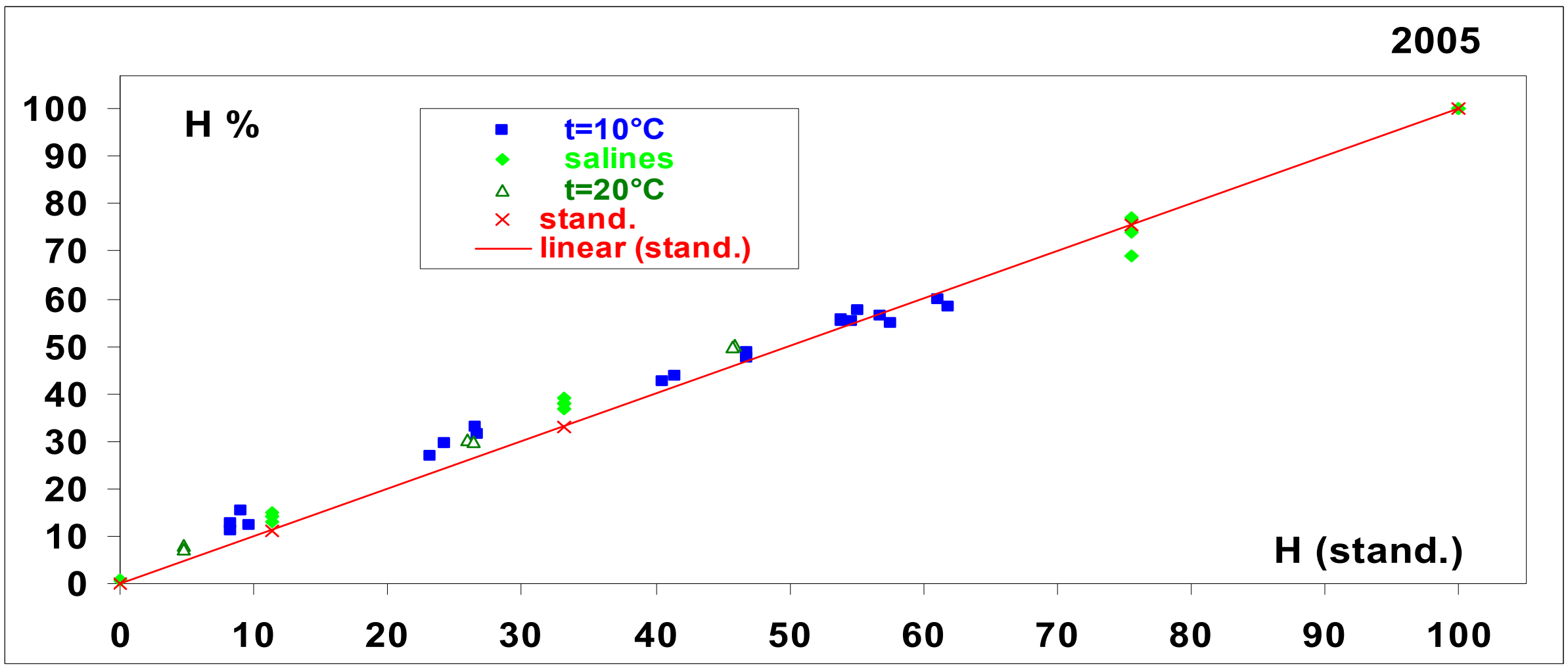

**Fig.** 7**b.** Relative humidity H(%) measured above saline solutions and in the barocamera with different pressure and temperature values, compared to the data obtained with the standard TOROS hygrometer and to table values for saline solutions.

**Figure** 7**b** indicates that the data obtained with our sensors in saline solutions for 20$^o$C and the atmospheric pressure, and those obtained in the barocamera with various temperature and pressure values, are mutually consistent and may be approximated with the same curve. This makes it possible to assume that the readings of the sensors are independent of pressure and temperature.

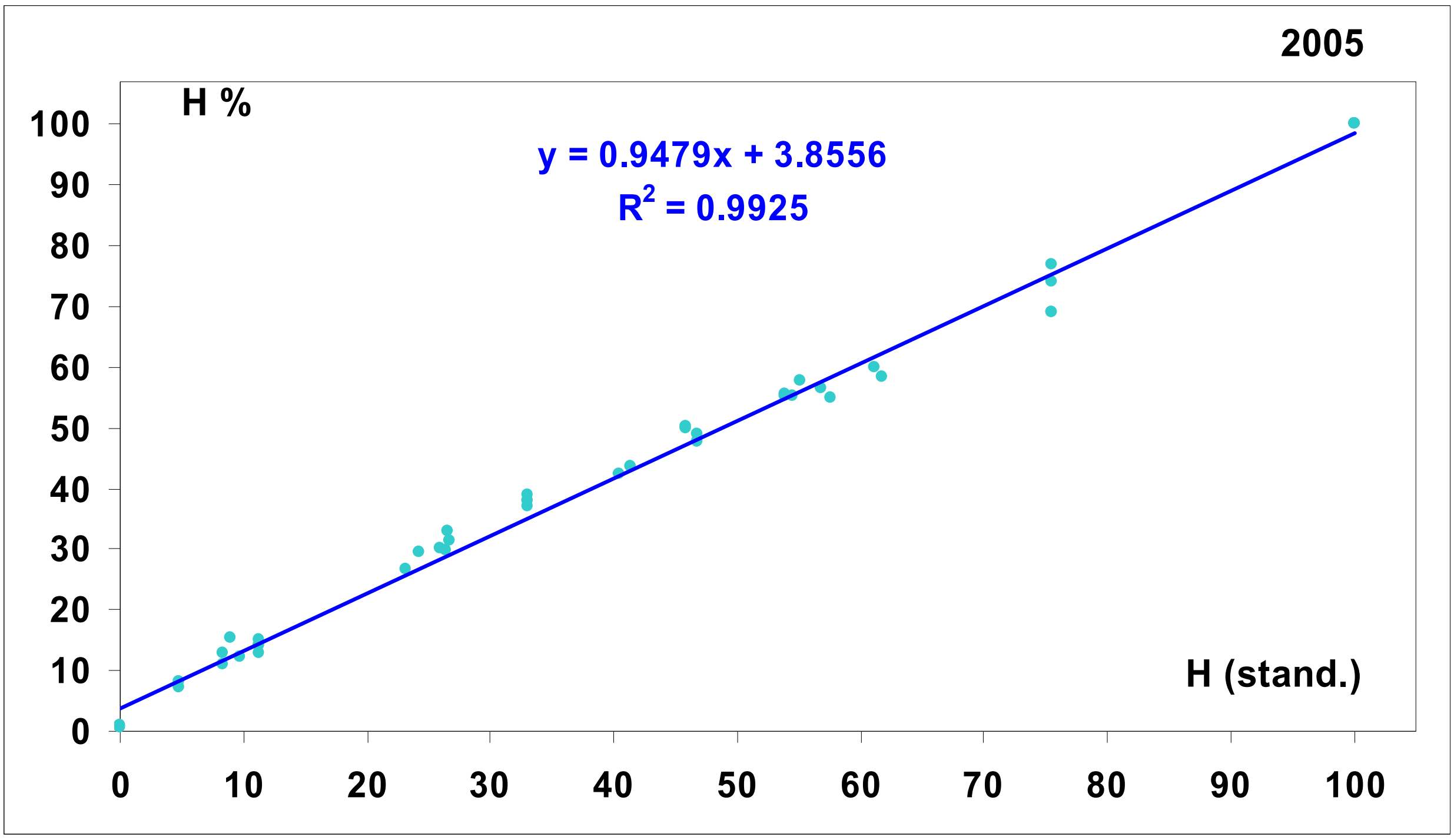

**Fig.7c.** Linear approximation of all data for all sensors for the measurements made in 2005.

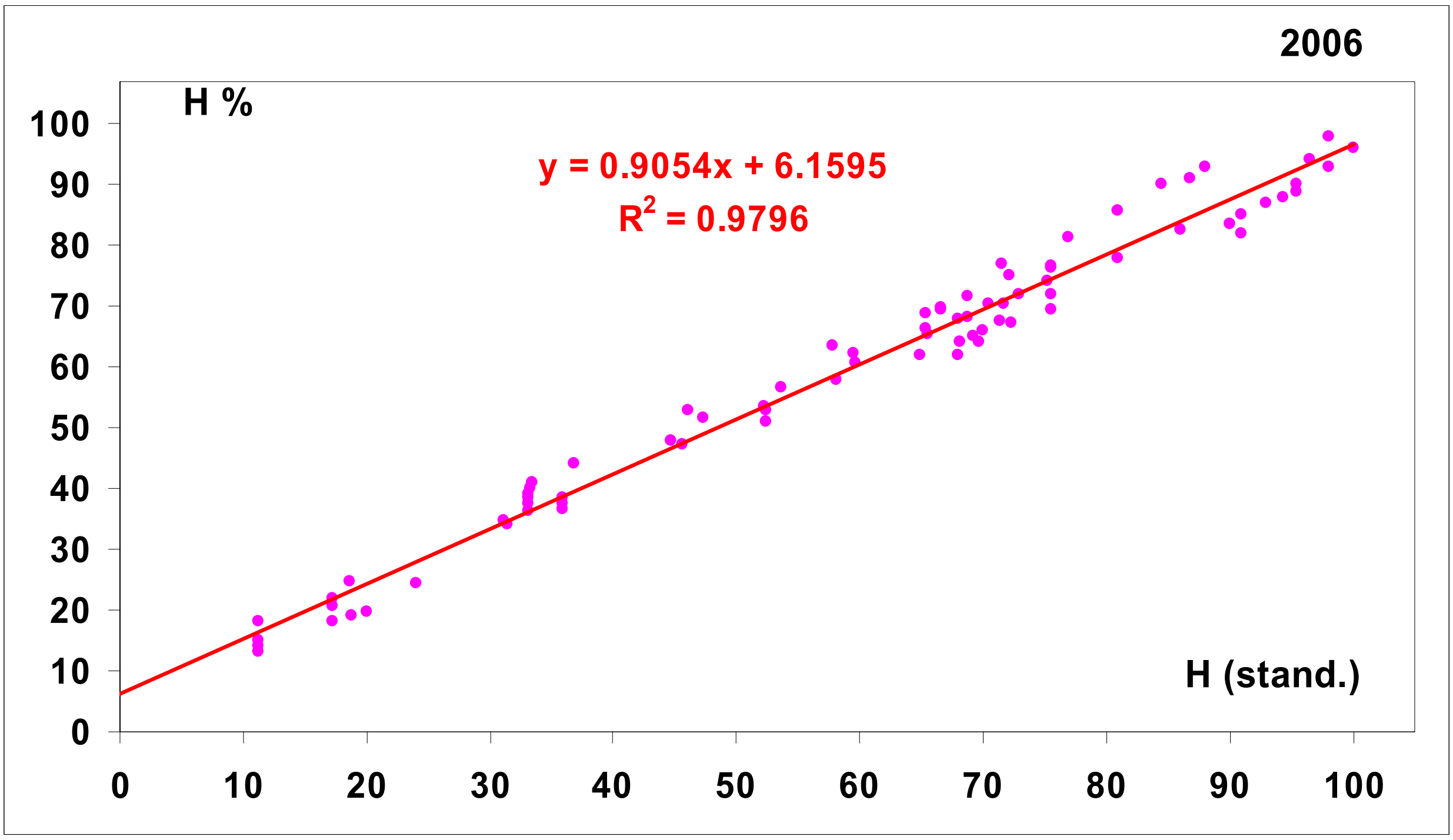

**Fig.7d.** Linear approximation of all data for all sensors for the measurements made in 2006.

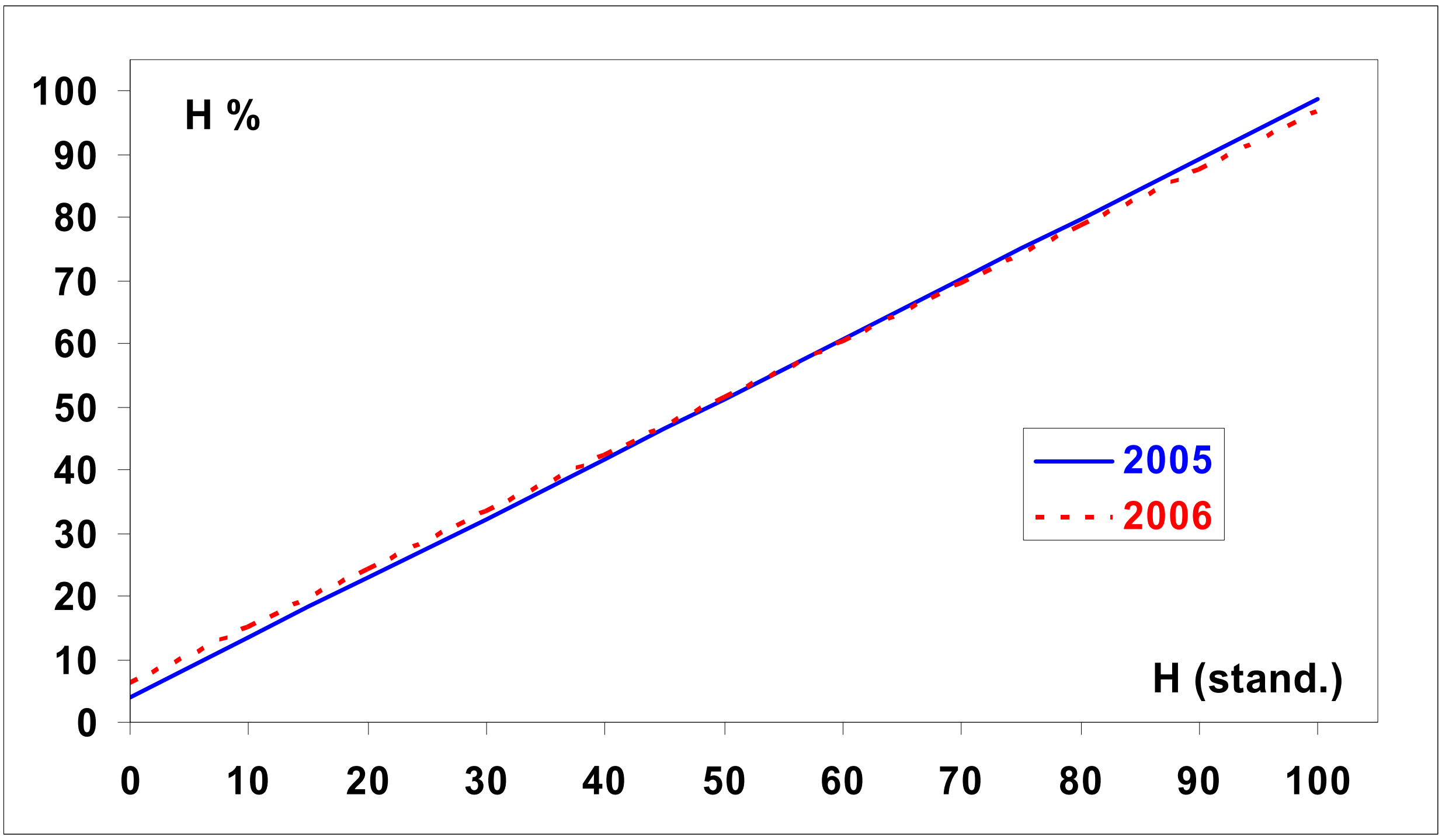

**Fig.8.** Linear approximations of the data for all sensors for the years 2005 and 2006.

**Figures 7c,d** presents linear approximation of all measurements made in 2005-6, while **Fig.8** displays the comparison of the linear approximations obtained in the years 2005 and 2006. The approximations for the years 2005 and 2006 are consistent within 2%, which provides evidence for stability of the measuring parameters of the sensors. However, the 2% measurement error for relative humidity along the total humidity scale, claimed by the producer in the passport of the sensor, indicates that measurements of absolute humidity for low values of relative humidity may display a large relative error.

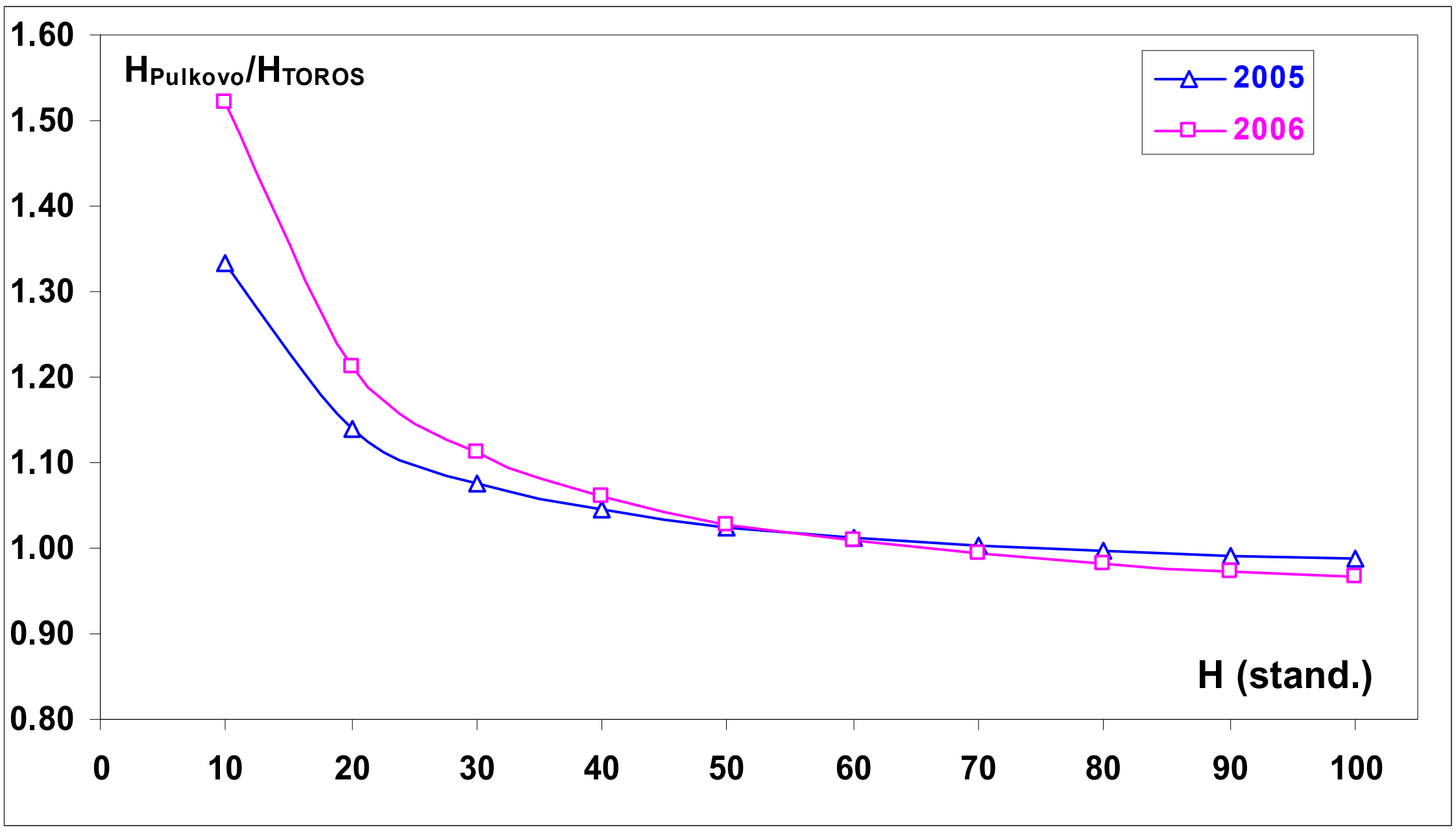

**Fig. 9**. The ratio of the Pulkovo and standard scale of humidity.

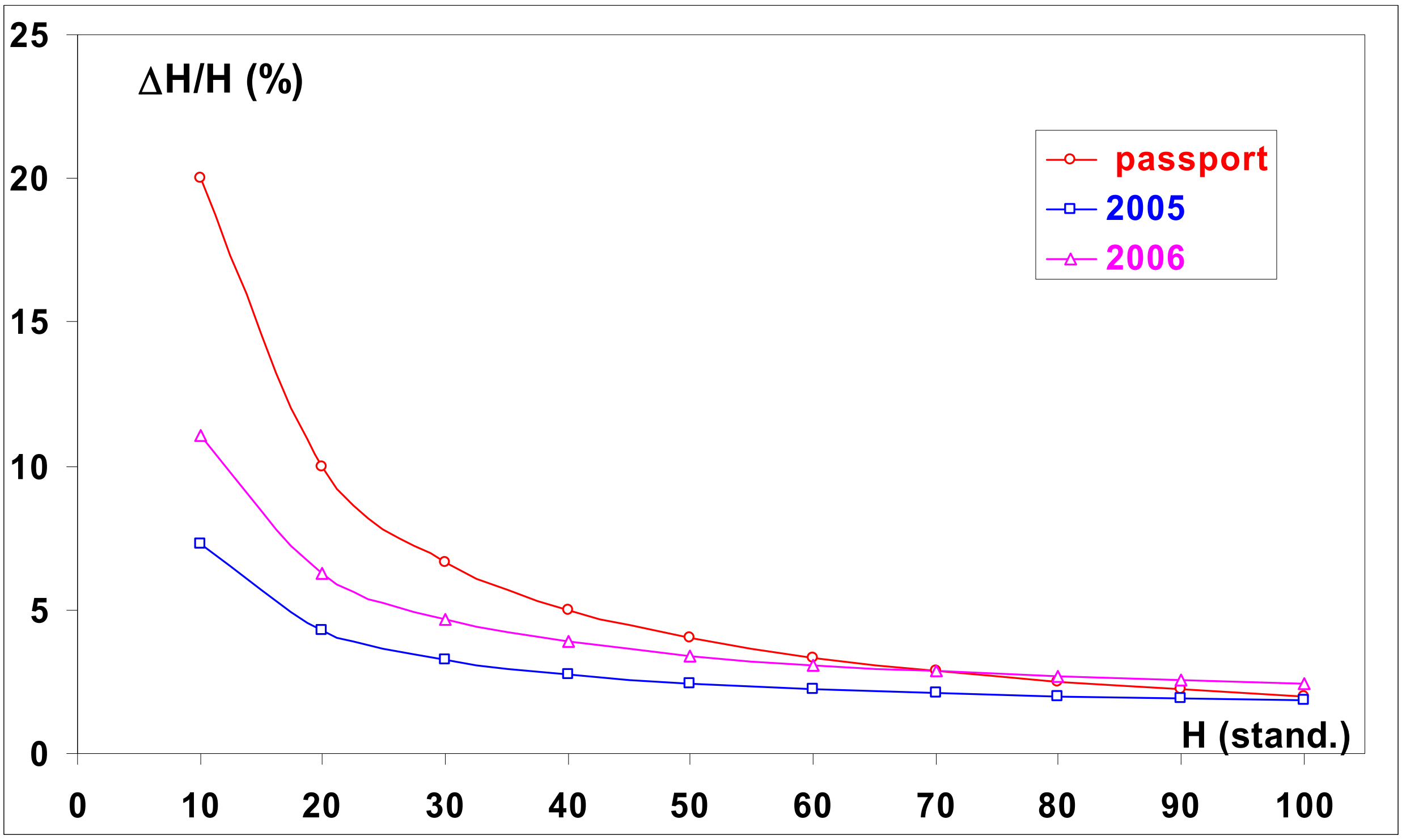

**Fig.10**. The comparison between the passport determination error for relative humidity and the errors derived from the calibrations in the years 2005 and 2006.

**Figure 9** presents the ratio of the relative humidity H(%) calculated from approximations that characterize the Pulkovo sensors and the standard relative humidity, i.e. that corresponds to the measurements made with standard equipment. **Figure 10** displays relative errors for H(%) derived from the errors of the approximation coefficients (see **Fig.8**) (for the years 2005 and 2006). The same Figure presents the relative error H(%) calculated according to the 2% passport measurement error for H(%) along the total scale of humidity. We can see from **Figs 9** and **10** that for low relative humidity, both relative and systematic error of determination of humidity increase, i.e. a systematic deviation of measured values of humidity from standard appears. This means that measurements made with low relative humidity may be burden with significant individual errors. Indeed, as we see from **Fig. 3**, measurements made with low pressure (which corresponds to small relative humidity) display substantial deviations of individual measurements from the average values. Thereby, we may conclude that measurements made with low relative humidity are hardly advisable when high accuracy of determination of the water vapor content in the cell is needed.

We attempted to consider individual parameters of our sensors in more detail and to study the possibility to reduce the measurement error by individual calibration of the sensors, using our measurements made in 2006, which present a larger volume of uniform data. **Figure 11** presents the results of the comparison between the readings of the sensor №1229 for two temperature values ($10^{o}$ and $20^{o}$C) and the standard values obtained in the measurements with saline solutions. The Figure displays a minor, within the errors, but clearly seen a difference between the approximations of measurements for these temperature values. Similar differences in calibrations for two temperature values may be also noticed for other sensors. Taking into account the fact that the temperature differences are within the error limits, individual relations for the transition from readings of Pulkovo sensors to standard values were derived for each sensor from the total set of data obtained with this sensor. Using the above relations, the measurements made in the cell were corrected, which, regrettably, did not result in any substantial improving of the convergence of the measured relative humidity values. The reason for that became clear after the calculation of individual correction errors for each sensor (**Fig. 12**). The correction error is not smaller than the measurement error, which does not make it possible to expect for any noticeable gain form the correction. Note that the problem of correction of measurements with the use of individual calibration for each sensor is justified. Essentially, it is reduced to providing the calibration for each sensor with the error smaller than the measurement error, i.e. to carrying out a sufficient number of measurements in order to construct a calibration curve.

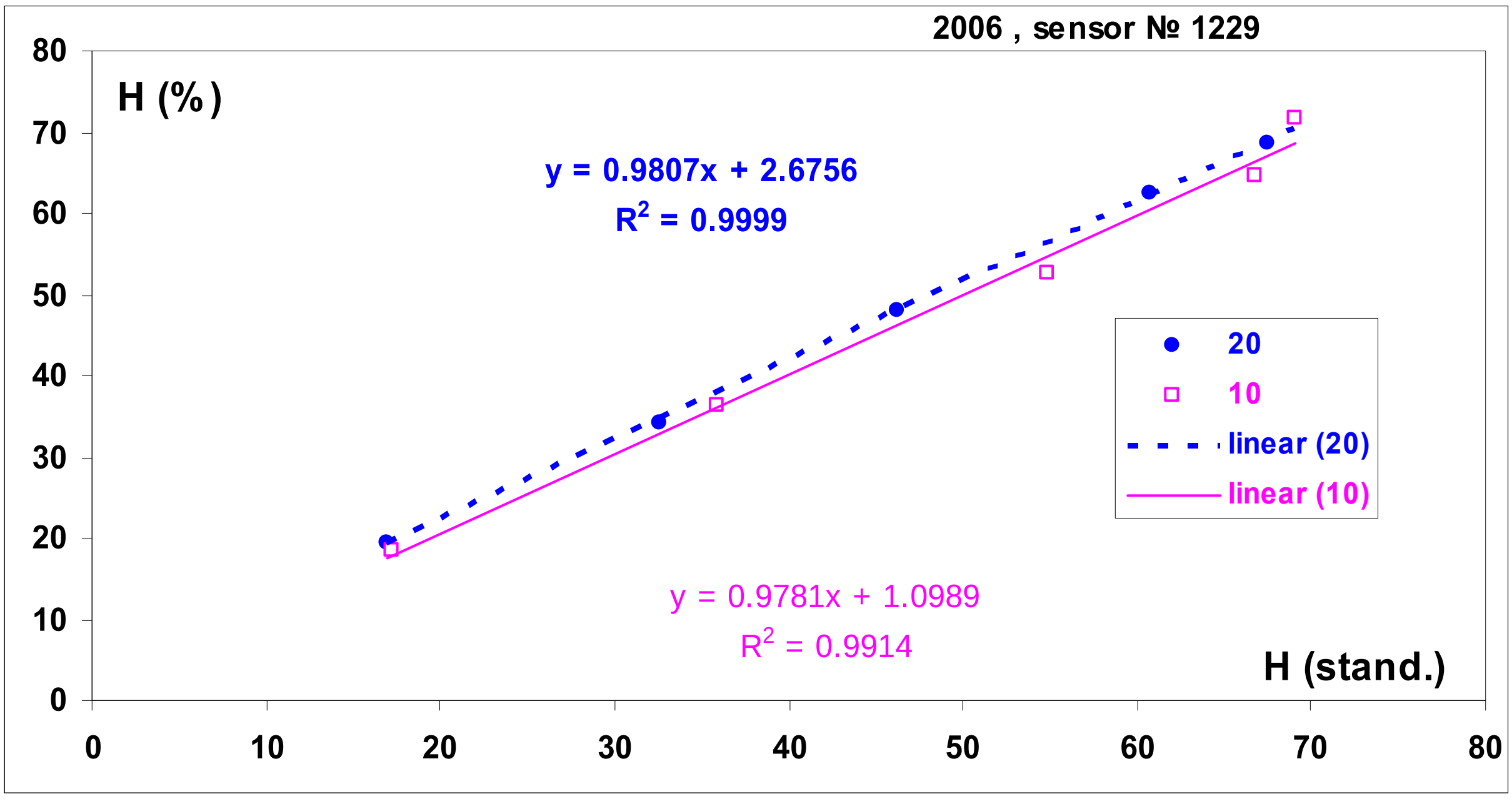

**Fig. 11**. A comparison between the data obtained for two temperature values.

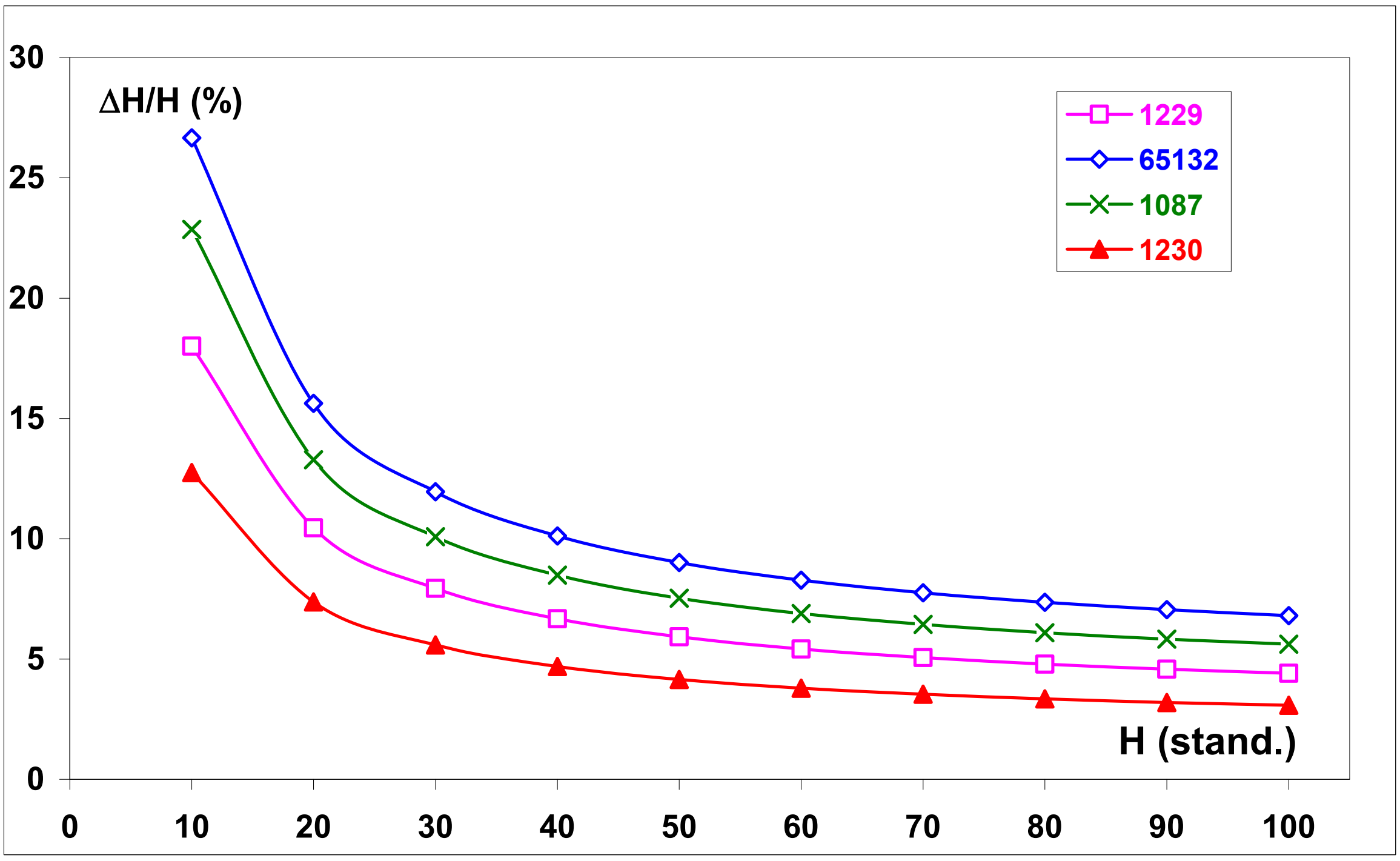

**Fig. 12**. The error of determination for H(%) (the correction error) calculated from the comparison between H(%) obtained with a given sensors and the TOROS hygrometer.

## Conclusion.

The analysis of the humidity measurements made in the Pulkovo vacuum cell in the years 2004 to 2006 with the use of sensors, and also testing of the sensors carried out in the PTU laboratory in Lindenberg, make it possible to conclude that:

- sensors of this type may be used for measurements and control of humidity in the cell;
- the sensor readings do not depend on air pressure;
- the results of the measurements with the sensors were stable over the time of their use (2004—2006);
- individual calibrations of the sensors showed that, within the passport accuracy, their measurement errors differ; among the examined sensors, №1229 and 1230 appeared to be the best;
- all the sensors displayed weak temperature dependence in the transition from 10$^{o}$C to 20 $^{o}$C, with a smaller measurement error for 20 $^{o}$C;
- the use of individual calibrations of the sensors for correction of measurements made with individual sensors did not yield the expected effect. This is explained by the limited volume of the measurements for individual calibrations. Further on, it is advisable to use such correction with more thorough calibrations;
- the analysis of individual calibrations and errors made it possible to discriminate the interval of the relative humidity, in which the optimum accuracy of determination of the water vapor content in the cell (H=40-80%) may be attained.

The testing of the sensors guarantees the determination of absolute humidity in the cell within the interval of relative humidity 40--80%, with the error of the order of 5% in the case of constant temperature along the cell, and 5--10% in the presence of temperature gradient and non-uniform distribution of water vapor along the cell, caused by introduction of liquid water. More detailed individual calibration of sensors will undoubtedly result in correction of the data of the measurements and an increase in the accuracy of determination of the water vapor content in the cell.

## References

1. **Leiterer U. et al.** Contr. to Atmosh. Phys., 1998, 71/4, pp. 401-420.
2. **Alekseeva G.A. et al.** International Conference ENVIROMIS 2000, Proceeding, ed. by E. Gordov, 2001, pp.38-42, Tomsk, (Environmental Observations, Modeling and Information Systems as Tools for Urban/Regional Pollution Reabilitation, Tomsk, Russia, 24-28 October 2000).
3. **Galkin V., Sal'nikov I., Naebert T., Nikanorova I., Leiterer U., Naebert T., Alekseeva G., Novikov V., Ilyin G., and Pakhomov V.**, Izv. GAO (rus), 2004, No. 217, pp. 472-484.
4. **Passport and service manual for IVTM-7 MK-C device for humidity and temperature measurements**. TFAP2.844.009, (rus), Moskow, OPEN COMPANY "PRAKTIC-NC".
5. **L. Greenspan**. Journal of Research by the National Bureau of Standards, 1977, 81A, pp. 89-96.
6. **Leiterer U. et al.** Contr. Atmosph. Phys.1997, 70, pp. 319-336.